\documentclass[twocolumn,aps,prb,amsmath,amssymb]{revtex4-1}
\usepackage[latin9]{inputenc}
\usepackage{color}
\usepackage{amsmath}
\usepackage{graphicx}

\makeatletter

\usepackage{dcolumn}
\usepackage{bm}
\usepackage{mathrsfs}

\makeatother

\begin{document}

\title{The effect of microwaves on superconductors for kinetic inductance
detection and parametric amplification}

\author{A.\,V.~Semenov}
\email[]{alexandre.semenov@gmail.com}

\affiliation{Physics Department, Moscow State University of Education, 1 Malaya
Pirogovskaya st., Moscow 119992, Russia}

\affiliation{Moscow Institute of Physics and Technology, Dolgoprudny, Moscow 141700,
Russia}

\author{ I.\,A.~Devyatov}
\email[]{Deceased on June 11th, 2019}

\affiliation{Lomonosov Moscow State University, Skobeltsyn Institute of Nuclear
Physics, 1(2), Leninskie gory, GSP-1, Moscow 119991, Russia}

\affiliation{Moscow Institute of Physics and Technology, Dolgoprudny, Moscow 141700,
Russia}

\author{M. P. Westig}

\affiliation{Kavli Institute of NanoScience, Faculty of Applied Sciences, Delft
University of Technology, Lorentzweg 1, 2628 CJ Delft, The Netherlands}

\author{T. M. Klapwijk}

\affiliation{Kavli Institute of NanoScience, Faculty of Applied Sciences, Delft
University of Technology, Lorentzweg 1, 2628 CJ Delft, The Netherlands}

\affiliation{Physics Department, Moscow State University of Education, 1 Malaya
Pirogovskaya st., Moscow 119992, Russia}


\date{\today}

\begin{abstract}
We address, using concepts of the microscopic theory of superconductivity,
parametric amplifiers and kinetic inductance detectors focusing on
the interaction of microwave radiation with the superconducting condensate.
This interaction was identified recently as the source of the apparent
dissipation in microwave superconducting micro-resonators at low temperatures.
Since the evaluation of the performance of practical devices based
only on the microwave-response is not sufficiently informative about
the underlying physical processes, we propose an experimental system
to measure the microscopically relevant spectral functions as well
as the non-equilibrium distribution function of a microwave-driven
superconducting wire. The results indicate the limits of the commonly
used phenomenological theories, providing the groundwork for further
optimisation of the performance. 
\end{abstract}
\maketitle
\section{Introduction}

In recent years there has been an increased interest in the use of
conventional superconductors in the presence of a microwave field,
for example in quantum computation \cite{Blais2004,Wallraff2004},
parametric amplification\cite{eom}, and for astronomical muliti-pixel
detection with microwave kinetic inductance sensors \cite{Zmuidzinas2012,dvnatcom}
The subject is also closely related to efforts to measure the Higgs
mode in superconductors \cite{Moor2017,Matsunaga2013,Matsunaga2014,Beck2013}.
The experiments are carried out far below the critical temperature
of the superconductor, where few quasi-particles are present and the
properties in response to the microwave field are dominated by the
superconducting condensate.

A commonly used assumption for the non-linear response of a superconductor
is summarized in writing the kinetic inductance as 
\begin{equation}
L_{k}(I)\approx L_{k}(0)\left[1+\left(\frac{I}{I_{*}}\right)^{2}\right]\label{NonlinearInductance}
\end{equation}
with $I_{*}$ the scale of the non-linearity and $L_{k}(0)={\hbar R_{n}/{\pi\Delta}}$,
with $R_{n}$ the resistance in the normal state and $\Delta$ the
superconducting energy gap. Such an expression is the adaptation of
the standard Ginzburg-Landau analysis of a dc current-carrying superconductor\cite{Tinkham},
assuming $I/{I_{*}}\ll1$. The underlying microscopic picture is a
supercurrent carried by Cooper-pairs, which in rest have net zero
momentum $(\vec{k}\uparrow,-\vec{k}\downarrow)$. When a supercurrent
flows all Cooper-pairs have a net momentum, $\vec{p}{_{s}}$, or pairs
with $((\vec{k}+\vec{p}{_{s}})\uparrow,(-\vec{k}+\vec{p}{_{s}})\downarrow)$.
The kinetic energy stored in the moving condensate goes at the expense
of the net condensation energy, which results in a reduced order parameter
$\Psi$, {\emph{i}.e.} a reduced energy gap $\Delta$. In order
to apply this analysis to the kinetic inductance at high frequencies,
ranging from microwave to THz frequencies, the time-response of the
system is important too\cite{Anlage}. For an instantaneous response
of the order parameter to the change in the supercurrent the quantity
$I_{*}$ differs from the one with a delayed response, because of
the very long relaxation time. Such a time-delay has been experimentally
observed\cite{Tinkham} by applying current-pulses with a current
larger than the critical current, causing, with indium as a superconductor,
a time-delay in the order of nanoseconds in agreement with an energy-relaxation
time of 148 psec. Both the Ginzburg-Landau analysis as well as the
analytical expressions for the non-equilibrium response are applicable
only close to the critical temperature of the superconductor, $T_{c}$.
The new applications are at much lower temperatures, where the order
parameter is energy-dependent, and the response to radiation needs
to take into account the change in the density-of-states (DOS) due
to the absorbed radiation. In addition, the microwave-frequency in
comparison to characteristic relaxation times needs to be considered.

Experimentally, it was demonstrated by De Visser \textit{et al}\cite{dv},
that the resonant frequency of an Al superconducting resonator shifts
with increased microwave power. This shift appears to be analogous
to a temperature rise although the applied frequency $\omega_{0}$
has a photon-energy much lower than the energy gap $\Delta$, which
rules out pair-breaking by the photon-energy. In addition, because
the measurements are carried out far below $T_{c}$ the density of
quasiparticles is very low. Therefore, we studied theoretically the
nature of a superconducting condensate, which oscillates at a frequency
$\omega_{0}$ due to an applied microwave field. It was demonstrated\cite{semprl}
that the microwave field has a depairing effect on the superconductor,
analogous to that of a dc current\cite{bar,kupr1,rom,ant}, but qualitatively
different. The DOS loses the sharp peak at the gap energy, which is
comparable to what happens with a dc current, but in addition it acquires
features at specific energies $\Delta\pm n\hbar\omega_{0}$, with
$\Delta$ the modulus of the order parameter. These features in the
density-of-states are a manifestation of Floquet states, which are
the eigenstates of any quantum-system, exposed to a periodic field
\cite{grif}. 
It was also shown \cite{semprl} that the DOS develops an exponential-like
tail in the sub-gap region.

The present study is carried out to relate the phenomenological expression
of Eq. \ref{NonlinearInductance} and the recently developed microscopic
properties of a superconductor with an oscillating condensate. On
general grounds we expect that the quadratic dependence does not change,
but we like to be able to calculate the parameters. In addition, the
conceptual understanding of the response of a uniform superconductor,
such as aluminium, assumed here, may provide insight to understand
also the difference with the response of inhomogeneous superconductors
such as niobium titanium nitride (NbTiN)\cite{Driessen2012} or granular
aluminium (GrAl)\cite{Gruenhaupt2019}. In the present article we
present additional theoretical results for a realistic case by including
inelastic scattering. In addition, we present the design of an experiment
which would enable a measurement of the microscopic parameters with
a tunnel-probe of a superconductor exposed to a microwave-field, while
at the same time avoiding that the tunnel-process, intended as a passive
probe, is effected by the microwave field.

\section{Action of microwaves on the superconducting condensate}

In order to go beyond the phenomenological Ginzburg-Landau theory
we need to use the microscopic theory of non-equilibrium superconductivity
\cite{laov,bel,ram,Kopnin}. It allows us to access the practically
relevant regime of $\Delta\gg k_{B}T$ and it includes the fact that
the superconducting properties are dependent on the energy. This dependence
is very well known from tunnelling experiments, but it also enters
the response of the superconducting condensate to microwave radiation.
We assume a dirty superconductor \emph{i.e.} with an elastic mean
free path $\ell$ much smaller than the BCS coherence length $\xi_{0}$,
meaning that we can rely on the Usadel-theory\cite{Usadel} for impurity-averaged
Green's functions. As shown by Stoof and Nazarov\cite{Stoof1996},
for the experimental conditions met in the present subject the theory
can conveniently be expressed in the complex function $\theta(E)$
and the real function $\phi(E)$.

The retarded and advanced Green's functions are expressed geometrically
by two matrices 
\[
\hat{G}^{R}=\left({\begin{array}{cc}
\cos\theta & e^{-i\phi}\sin\theta\\
e^{i\phi}\sin\theta & -\cos\theta
\end{array}}\right)
\]

and

\[
\hat{G}^{A}=\left({\begin{array}{cc}
-\cos\bar{\theta} & e^{-i\phi}\sin\bar{\theta}\\
e^{i\phi}\sin\bar{\theta} & -\cos\bar{\theta}
\end{array}}\right)
\]
with $\theta=\theta(r,E)$ a complex angle which is a measure of the
pairing, for short called the pairing angle, and $\phi=\phi(r,E)$
is the superconducting phase, a real quantity. With these variables
one expresses quantities familiar from the Ginzburg-Landau theory
such as the supercurrent $J_{s}$ and the density of superconducting
electrons $|\Psi|^{2}$ in microscopic variables. For the supercurrent
we have:

\begin{equation}
j_{s}=\frac{\sigma_{N}}{e}\int_{-\infty}^{+\infty}dE\tanh\left(\frac{E}{2k_{B}T}\right)\mathrm{Im}\sin^{2}\theta\left(\nabla\phi-\frac{2e}{\hbar}\vec{A}\right)\label{supercurrent}
\end{equation}
and for the density of superconducting electrons:

\begin{equation}
|\Psi|^{2}=\frac{m}{e^{2}\hbar}\sigma_{N}\int_{0}^{+\infty}dE\tanh\left(\frac{E}{2k_{B}T}\right)\mathrm{Im}\sin^{2}\theta\label{superelectrondensity}
\end{equation}
Here, $\sigma_{N}=e^{2}N_{0}D$ is the normal state
conductivity, with $N_{0}$ the density-of-states in the normal state, including  spin,
and $D$ the diffusion coefficient. $m$ is the electron mass. The
second quantity makes clear that the density of superconducting electrons
is determined by $\mathrm{Im}[\sin^{2}\theta]$, which is equivalent
to an effective energy dependent density of pairs. The integration
over the energies weighted with the Fermi-Dirac distribution determines
the averaged quantity $|\Psi|^{2}$. So for a proper understanding
of the response of the superconductor one needs to know $\theta(E)$
and $\phi(E)$. The kinetic inductance is determined by the density
of superconducting electrons through:

\begin{equation}
L_{k}=\frac{m}{e}\frac{1}{e\sigma_{N}|\Psi|^{2}}\label{kinind}
\end{equation}
which illustrates that the non-linear response of the kinetic inductance
is due to a change of the density of superconducting electrons, which
on its turn is determined by the energy-dependent pairing angle $\theta$.
The single particle density-of-states, which is the quantity which
is measured with a tunnel-junction, is given by 
\begin{equation}
N(r,E)=N_{0}\mathrm{Re}{\cos\theta(r,E)}.\label{tunn}
\end{equation}

\subsection{DC currents and microwave currents}

For a stationary current-carrying superconductor Anthore \emph{et
al}\cite{ant} have shown that the quantities $\theta$ and $\phi$
are determined by two basic equations:

\begin{equation}
E+i\Gamma\cos\theta=i\Delta\frac{\cos\theta}{\sin\theta}\label{dccurrent1}
\end{equation}

and 
\begin{equation}
\vec{\nabla}(\vec{v}_{s}\sin^{2}\theta)=0\label{dccurrent2}
\end{equation}
with $\vec{v}_{s}=D[\vec{\nabla}\phi-({2e}/\hbar)\vec{A}]$ and $\Gamma$
given by $\left(\hbar/2D\right)v_{s}^{2}$. Experimentally, either
a magnetic field or a current is imposed forcing a value for $\Gamma$,
which then leads to solutions of Eq.~\eqref{dccurrent1}, i.e. for
$\theta(E)$. For later use we rewrite Eq.~\eqref{dccurrent1} to
\begin{equation}
iE\sin\theta+\Delta\cos\theta+\alpha_{dc}\Pi=0\label{theta_dccurrent}
\end{equation}
with $\alpha_{dc}$ defined as $\Gamma/4$ and $\Pi=4i\cos\theta\sin\theta=2i\sin\theta$.
For $\alpha_{dc}=0$ we find the conventional BCS-solution. With finite
$\alpha_{dc}$ the BCS density-of-states is rounded as well as a reduced
value for the energy gap in the excitation-spectrum is obtained. We
assume here a uniform current over the cross-section of the wire.

The effect of a dc supercurrent and a magnetic field on the superconducting
state has been measured by Anthore \emph{et al} \cite{ant}, \emph{i.e.}
the effect on $\theta(r,E)$ by measuring the density-of-states of
the superconductor with a tunnel-junction. The results illustrate
that for a uniform current density and for a narrow strip in a magnetic
field the response of the superconductor is for low current densities
identical. For higher values of $\alpha$ a difference occurs when
the supercurrent reaches the critical pair-breaking current at which
point stable solutions seize to exist. For the magnetic field solutions
continue to exist going down smoothly until a gapless state is reached.
The change in $\theta(r,E)$ enters also the kinetic inductance, through
Eqs.~\eqref{superelectrondensity} and \eqref{kinind} leading to
an increase in the kinetic inductance due to a reduction in the density
of superconducting electrons, which reflects a reduction in the pairing
angle $\theta(r,E)$. At small current, $I/{I_{*}}\ll1$, this increase
of kinetic inductance with the current is given by Eq. \eqref{NonlinearInductance}
with $I_{*}\simeq2.69I_{c}$, where the depairing critical current
$I_{c}\simeq0.75\Delta_{u}/eR_{\xi}$ \cite{ant}, $R_{\xi}$ is the
normal resistance per the coherence length $\xi=\sqrt{\hbar D/\Delta_{u}}$,
with $\Delta_{u}$ the unperturbed value of the energy gap. The zero-current
kinetic inductance, in the limit $\Delta_{u}\gg k_{B}T$, is given
by $L_{k}\left(0\right)=\hbar R_{N}/\pi\Delta_{u}$, with $R_{N}$
the normal resistance.

An electromagnetic field, defined as the vector potential $A$, represents
the microwave field $A=A_{0}\cos(\omega_{0}t)$ with frequency $\omega_{0}$,
which leads to an ac supercurrent. In using the Usadel equations,
we assume a dirty superconductor in which the momentum of the electrons
is randomised by impurity scattering faster than the relevant processes.
We will restrict the analysis to small intensities of the rf-drive
and frequencies less than the unperturbed energy gap $\Delta_{u}$:
\begin{equation}
\alpha\ll\hbar\omega_{0}\ll\Delta_{u},\label{lim1}
\end{equation}
with the parameter $\alpha$ the normalized intensity of the rf-drive
\cite{key-1}: $\alpha=e^{2}DA_{0}^{2}/4\hbar$. The inequalities
of Eq.~\eqref{lim1} impose the same restriction on $\alpha$, $\omega_{0}$
and $\Delta_{u}$ as used previously in Semenov et al\cite{semprl},
which means that the conditions for the 'quantum mode of depairing'
are fulfilled \cite{endnote1}.{} We assume that the temperature
is low, $k_{B}T\ll\Delta_{u}$, hence the number of thermal quasiparticles
at energies of order of $\Delta_{u}$ is negligible. While evaluating
the tunnel-relaxation model, we also assume that $\alpha\ll\Gamma_{inel}$,
which is a technical assumption required to apply linear expansion
of the Green functions in $\alpha$, and does not affect any of our
results qualitatively.

The response to an ac current with frequency $\omega_{0}$, in the
microwave-range, has been presented in Semenov et al\cite{semprl}
and rewritten in the variables $\theta$ and $\phi$ it leads again
to Eq.~\ref{theta_dccurrent} with $\alpha_{dc}\rightarrow\alpha$
and the function $\Pi$ replaced by: 
\begin{equation}
\begin{array}{c}
\Pi=i\sin\theta(\cos\theta_{+}+\cos\theta_{-})+i\cos\theta(\sin\theta_{+}+\sin\theta_{-})\end{array}\label{pi_ac}
\end{equation}
with the subscripts representing the argument being $E+\hbar\omega_{0}$
or $E-\hbar\omega_{0}$. It represents the moving superconducting
condensate due to the oscillating microwave currents.

To facilitate the comparison with experimentally more accessible values,
we express $\alpha/\Delta_{u}$ in terms of the induced rf supercurrent
$I_{rf}$. One can relate the amplitude of the induced current $I_{0}$
and the field as $L_{k,u}I_{0}=A_{0}$, where $L_{k,u}$ is defined
per unit length along the wire. This is just Eq.~\eqref{supercurrent}
in the limit of a small current density, without the phase gradient.
One arrives at 
\begin{equation}
\frac{\alpha}{\Delta_{u}}.=\frac{1}{2\pi^{2}}\frac{\left\langle I_{rf}^{2}\right\rangle }{\left(\Delta_{u}/eR_{\xi}\right)^{2}}\simeq0.028\frac{\left\langle I_{rf}^{2}\right\rangle }{I_{c}^{2}}.\label{alpha_to_I}
\end{equation}
by expressing $\alpha$ through $I_{rf}$ as $\alpha=e^{2}D\left(L_{k,u}I_{0}\right)^{2}/4\hbar=e^{2}D\left(L_{k,u}\right)^{2}\left\langle I_{rf}^{2}\right\rangle /2\hbar$,
with $\left\langle I_{rf}^{2}\right\rangle =I_{0}^{2}/2$ the mean
square of the induced rf-current.

\subsection{Inelastic scattering}

At any finite temperature, the presence of microwave
results in an absorption of microwave energy by electrons, which needs
to be balanced by an inelastic scattering process. Hence, in the kinetic equation
we take into account inelastic scattering, which for consistency should
also be studied for the spectral properties. In Semenov et al\cite{semprl}
this was not done explicitly with the assumption that the presence
of quasiparticle relaxation was used implicitly. Without a strong
enough relaxation the distribution function can not have the equilibrium
form assumed in the previous work. Inelastic scattering is introduced
by assuming a relaxation time approximation, which is equivalent to
take the self-energy of the form \cite{melkop,TSK}: 
\begin{equation}
\breve{\Sigma}_{inel}=-i\Gamma_{inel}\breve{G}_{res},\label{selfenergy}
\end{equation}
with $\Gamma_{inel}$ the tunneling rate and $\breve{G}_{res}$ the
Green's function of an equilibrium 'reservoir' to which the 'hot electrons'
tunnel. Formally, this model corresponds to the relaxation time approximation.
In principle, it can be compared to a thin film superconductor coupled
to a large normal reservoir \emph{via} a tunnel barrier with a transparency,
equivalent to a tunnel-rate equal to $\Gamma_{inel}$. While, under
realistic conditions, tunnel-coupling to a reservoir is not the mechanism
of energy relaxation, it is a very useful and tractable model, which
captures the essential physics. Its predictions about the effect of
the microwave drive on the spectral functions remains qualitatively
correct for the case of electron-electron or electron-phonon interaction
inside the superconductor. Moreover, as we will discuss below, the
effect of the ac-drive on quantities like the order parameter and
the kinetic inductance are insensitive to details of the superconductor's
spectral properties introduced by the inelastic processes. Hence,
the corresponding results derived with the chosen model are correct
quantitatively.

\begin{figure}[h]
\centering \includegraphics[width=1\columnwidth]{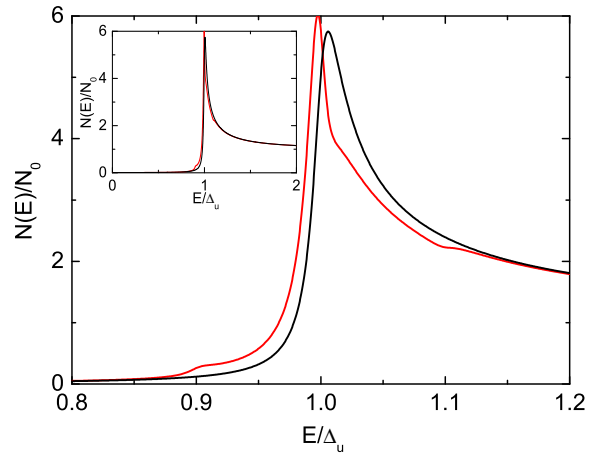} \caption{The normalized DOS of a superconductor $N(E)/N_{0}=Re[\cos\theta(E)]$
with $\alpha/\Delta_{u}=10^{-3}$ and $\hbar\omega_{0}/\Delta_{u}=0.1$
on an expanded energy-scale. The red curve is with radiation and the black
without. A total inelastic parameter is assumed of $\Gamma_{inel}/\Delta_{u}=0.01$.
The inset shows the full density of states making clear that the deviations
are small, but observable in order to evaluate the conceptual framework.}
\label{Figure1_DOS} 
\end{figure}

In terms of the pairing angle $\theta$ one obtains:

\begin{equation}
(iE-\Gamma_{inel})\sin\theta+\Delta\cos\theta+\alpha\Pi=0\label{theta_accurrent2_inel}
\end{equation}
This expression provides solutions for $\theta(E)$ for a given value
of $\alpha$ and for a material dependent inelastic scattering rate
$\Gamma_{inel}$. A typical result for the density-of-states, Eq.~\eqref{tunn},
is shown in Fig.~ \ref{Figure1_DOS} using results to be presented
in the next section.

\subsection{Analytical results}

Employing the self energy Eq.\eqref{selfenergy}, one comes to the
kinetic equation for the stationary longitudinal components \cite{laov,bel,ram}
of the quasiparticle distribution function $f_{L}$ :

\begin{equation}
I_{phot}\left[f_{L}\right]+I_{inel}\left[f_{L}\right]=0.\label{kin}
\end{equation}

Here, the electron-photon collision term $I_{phot}$ describes creation
of the quasiparticles and absorption of energy, and the inelastic
scattering term $I_{inel}$ provides quasiparticle and energy relaxation.
The integral of electron-photon collisions is given by:

\begin{equation}
I_{phot}=\alpha\left(R_{+}\left(f_{L}-f_{L+}\right)+R_{-}\left(f_{L}-f_{L-}\right)\right),\label{I_phot}
\end{equation}

\noindent with $R_{\pm}=\mathrm{Re}[\cos\theta_{\pm}]+{\mathrm{Im}[\sin\theta]\mathrm{Im}[\sin\theta_{\pm}}]/{\mathrm{Re}[\cos\theta}]$
(more details on the derivation of this electron-photon collision
integral can be found in Semenov \emph{et al}\cite{sem1}).

\noindent The integral of inelastic collisions in the relaxation time
approximation is

\begin{equation}
I_{inel}=\Gamma_{inel}\left(f_{L}-f_{L,res}\right)=\Gamma_{inel}\delta f_{L},\label{I_res}
\end{equation}
in which $f_{L,res}$ is the distribution function of quasiparticles
of the 'reservoir', which is assumed to be in equilibrium at a base
temperature $T$. The subscript L is used as a reminder that only
a longitudinal type of non-equilibrium, symmetric around $E_{F}$
is relevant\cite{Tinkham}.

The set of Eqs.~\eqref{theta_accurrent2_inel} and \eqref{kin} is
closed by the self-consistency equation, which has the usual form,

\begin{equation}
\Delta=\lambda\int\limits _{0}^{\hbar\omega_{D}}d\varepsilon f_{L}\mathrm{Im}\sin\theta,\label{delta}
\end{equation}

\noindent with $\omega_{D}$ the Debye frequency and $\lambda$ the
electron-phonon coupling constant.

The linearization of Eq.~\eqref{theta_accurrent2_inel} gives:

\begin{equation}
(E+i\Gamma_{inel})\delta\sin\theta-i\Delta_{u}\delta\cos\theta-i\cos\theta_{u}\delta\Delta+\alpha\Pi_{u}=0\label{R-Usadel}
\end{equation}
with $\delta\sin\theta\equiv\sin\theta-\sin\theta_{u}$, $\delta\cos\theta\equiv\cos\theta-\cos\theta_{u}$
. Here, $\theta_{u}$ denotes the unperturbed solution without rf-drive,
for $\alpha=0$, and is given by:

\begin{equation}
\cos\theta_{u}=\frac{\left(E+i\Gamma_{inel}\right)}{\left\{ \left(E+i\Gamma_{inel}\right)^{2}-\Delta_{u}^{2}\right\} ^{1/2}}\equiv\frac{\left(E+i\Gamma_{inel}\right)}{\Xi},\label{G0}
\end{equation}

\begin{equation}
i\sin\theta_{u}=\frac{-\Delta_{u}}{\left\{ \left(E+i\Gamma_{inel}\right)^{2}-\Delta_{u}^{2}\right\} ^{1/2}}\equiv-\frac{\Delta_{u}}{\Xi},\label{F0}
\end{equation}

\noindent with $\Delta_{u}$ the value of the order parameter for
no rf-drive and $\Xi\equiv\left\{ \left(E+i\Gamma_{inel}\right)^{2}-\Delta_{u}^{2}\right\} ^{1/2}$.
In the limit of $\Gamma_{inel}\rightarrow0$ the unperturbed functions
in Eqs.~\eqref{G0} and \eqref{F0} reduce to the standard BCS solution
\cite{laov,bel,ram}. The finite $\Gamma_{inel}$ describes the broadening
of the superconductor spectral functions due to the inelastic processes
\cite{naz}.

The solution of the linearized Eq.~\eqref{R-Usadel} has the form:

\begin{equation}
i\delta\sin\theta=i\frac{\partial\sin\theta}{\partial\Delta}_{|\alpha=0}\delta\Delta+i\frac{\partial\sin\theta}{\partial\alpha}_{|\Delta=\Delta_{u}}\alpha,\label{deltaF}
\end{equation}

\begin{equation}
\delta\cos\theta=\tan\theta_{u}\delta\sin\theta=\frac{\Delta_{u}}{E+i\Gamma_{inel}}i\delta\sin\theta,\label{deltaG}
\end{equation}

\noindent The partial derivatives are given by

\noindent 
\begin{equation}
\begin{aligned} & i\frac{\partial\sin\theta}{\partial\alpha}_{|\Delta=\Delta_{u}}=\\
 & =\frac{i\Delta_{u}\left\{ \left(E_{+}+i\Gamma_{inel}\right)+\left(E+i\Gamma_{inel}\right)\right\} \left(E+i\Gamma_{inel}\right)}{\Xi_{+}\Xi^{3}}+\\
 & +\left\{ E_{+}\rightarrow E_{-}\right\} ,
\end{aligned}
\label{partF}
\end{equation}

\noindent and

\begin{equation}
\begin{aligned}i\frac{\partial\sin\theta}{\partial\Delta}_{|\alpha=0}=\frac{-\left(E+i\Gamma_{inel}\right)^{2}}{\Xi^{3}}.\end{aligned}
\label{partF2}
\end{equation}

\noindent Eq.~\eqref{deltaF} expresses the linear change of the
Green function $\delta\sin\theta$ under the influence of rf-drive
as a sum of two terms: the one, proportional to the normalized rf-drive
intensity $\alpha$, and the other, proportional to the variation
of the order parameter $\delta\Delta$.

Since the change of the order parameter is determined in part by the
nonequilibrium distribution function of the quasiparticles, we first
determine this quantity from the kinetic equation. Just for simplicity,
here we restrict our derivations to the limit $k_{B}T\ll\hbar\omega_{0}$
(later, we will remove this restriction). Then the differences $f_{L,u}-f_{L\pm,u}$
are unequal to zero only in the small energy interval $-\hbar\omega_{0}<E<\hbar\omega_{0}$,
where $\mathrm{Re}\cos\theta_{u}\cong\mathrm{Re}\cos\theta_{\pm,0}\cong\Gamma_{inel}/\Delta_{u}$
and $\mathrm{Re}\sin\theta_{u}\cong\left({\Gamma_{inel}}/{\Delta_{u}}\right)\left(E/\Delta_{u}\right)\ll\mathrm{Re}\cos\theta_{u}$.
Hence, the electron-photon collision integral \eqref{I_phot} can
be simplified to $I_{phot,u}=\alpha\left({\Gamma_{inel}}/{\Delta_{u}}\right)\left\{ \left(f_{L,u}-f_{L+,u}\right)+\left(f_{L,u}-f_{L-,u}\right)\right\} $.
Then the solution of the kinetic equation \eqref{kin} has the following
form:

\begin{equation}
\delta f_{L}=\frac{\partial f_{L}}{\partial\alpha}_{|\Delta=\Delta_{u}}\alpha,\label{deltafl3}
\end{equation}

\noindent with

\begin{equation}
\frac{\partial f_{L}}{\partial\alpha}_{|\Delta=\Delta_{u}}\begin{array}{c}
=\left\{ \begin{array}{c}
-\frac{2}{\Delta_{u}},\:E\in(0,\hbar\omega_{0})\\
\frac{2}{\Delta_{u}},\:E\in(-\hbar\omega_{0},0)\\
0,\:E\notin(-\hbar\omega_{0},\hbar\omega_{0})
\end{array}\right\} .\end{array}\label{deltafl4}
\end{equation}

\noindent Note that, with the chosen model of relaxation \eqref{selfenergy},
the quantity, which characterizes the strength of the inelastic interaction,
$\Gamma_{inel}$, drops out of the answer. The reason is that both
collision integrals in the kinetic equation \eqref{kin} are proportional
to $\Gamma_{inel}$. 

The linearization of the self-consistency equation \eqref{delta}
leads to the following relation for the small correction of the order
parameter $\delta\Delta$:

\begin{equation}
\begin{aligned}\delta\Delta=\delta_{\theta}\Delta+\delta_{f_{L}}\Delta.\end{aligned}
\label{vardel}
\end{equation}

\noindent with

\begin{equation}
\begin{aligned}\delta_{\theta}\Delta=\alpha\int_{0}^{\infty}dE\frac{\partial\mathrm{Im}[{\sin\theta}]}{\partial\alpha}_{|\Delta=\Delta_{u}}f_{L,u},\end{aligned}
\label{delF}
\end{equation}

\noindent and

\begin{equation}
\delta_{f_{L}}\Delta=\alpha\int_{0}^{\infty}dE\mathrm{Im}[{\sin\theta_{u}]}\frac{\partial f_{L}}{\partial\alpha}_{|\Delta=\Delta_{u}}\label{delfl}
\end{equation}

\noindent Here, Eq.~\eqref{delF} describes the change of the order
parameter $\Delta$ due to the change of the anomalous Green function
$\sin\theta$, and Eq.~\eqref{delfl} - due to the change of the
distribution function $f_{L}$.

Substituting into Eq.~\eqref{delF} and Eq.~\eqref{delfl} the formulas
for $\partial_{\alpha}{\sin\theta}_{|\Delta=\Delta_{u}}$ Eq.~\eqref{partF}
and $\partial_{\alpha}{f_{L}}_{|\Delta=\Delta_{u}}$ Eq.~\eqref{deltafl4}
makes it possible to calculate these corrections analytically. Taking
into account that the integrand in Eq.~\eqref{delfl} is analytical
in the upper half-part of the complex plane and decays faster than
$1/E$ at infinity, we replace the integration over the real semi-axis
$E\in\left(0,+\infty\right)$ to the integration over the imaginary
semi-axis $iE\in\left(0,+i\infty\right)$ and obtain, after dropping
terms of nonzero order in ${\Gamma_{inel}}/{\Delta_{u}}$ and ${\hbar\omega_{0}}/{\Delta_{u}}$:

\begin{equation}
\frac{\delta_{\theta}\Delta}{\alpha}\cong\int_{0}^{\infty}dy\frac{4y^{2}}{\left\{ y^{2}+1\right\} ^{2}}=-\pi.
\end{equation}
The correction of $\Delta$ due to the change of $f_{L}$ turns out
to be small:

\begin{equation}
\frac{\delta_{f_{L}}\Delta}{\alpha}\cong-\frac{\Gamma_{inel}}{\Delta_{u}}\int_{0}^{\frac{\hbar\omega_{0}}{\Delta_{u}}}xdx\simeq-\left(\frac{\hbar\omega_{0}}{\Delta_{u}}\right)^{2}\frac{\Gamma_{inel}}{\Delta_{u}},
\end{equation}

\noindent and can be neglected compared to $\delta_{F}\Delta$. Finally,
we obtain:

\begin{equation}
\frac{\delta\Delta}{\Delta_{u}}\cong-\pi\frac{\alpha}{\Delta_{u}}.\label{delta_Delta}
\end{equation}
The change in the distribution function $\delta f_{L}$, given by
Eqs.~\eqref{deltafl3}, \eqref{deltafl4} has a minor effect on $\delta\Delta$
because $\delta f_{L}$ is nonzero only in the small energy interval
$E\in(-\hbar\omega_{0},\hbar\omega_{0})$, where $\mathrm{Im}{\sin\theta_{u}}$
is small. It is obvious that the same holds if the temperature is
not small compared to $\hbar\omega_{0}$ (but still small compared
to $\Delta_{u}$). In the case $\hbar\omega_{0}\ll k_{B}T\ll\Delta_{u}$,
$\delta f_{L}$ is nonzero roughly at $E\in(-k_{B}T,k_{B}T)$. Hence
in the formula for $\delta_{f_{L}}\Delta$ one has to replace $\left(\hbar\omega_{0}/\Delta_{u}\right)^{2}$
by approximately $\left(k_{B}T/\Delta_{u}\right)^{2}$, which is also
a small factor .

The main results we obtain with this simplest possible model of inelastic
processes, are {i)} the formulas for the rf-drive induced corrections
to spectral functions and to the order parameter Eq.~\eqref{deltaF}
and Eq.~\eqref{delta_Delta}, and {ii)} the statement about smallness
of the effect of the rf-drive induced nonequlibrium in the quasiparticle
subsystem on the spectral functions.

\section{Superconducting density-of-states and microwaves}

The solution of Eqs.~\eqref{theta_accurrent2_inel} and \eqref{delta}
provides the superconducting properties expressed in $\theta\left(E\right)$,
which is dependent on the microwave-frequency $\omega_{0}$ and its
intensity, $\alpha$. The most direct manifestation of this change
due to the embedded microwave field, is the change of DOS, as defined
in Eq.~\eqref{tunn}. Previously, we have presented \cite{semprl}
results of the modified DOS for the case without inelastic processes,
which physically corresponds to $\varGamma_{inel}\ll\alpha$. Here,
we expand on those results by also calculating the change of DOS for
the opposite case $\varGamma_{inel}>\alpha$.

\begin{figure}[h]
\centering \includegraphics[width=1\columnwidth]{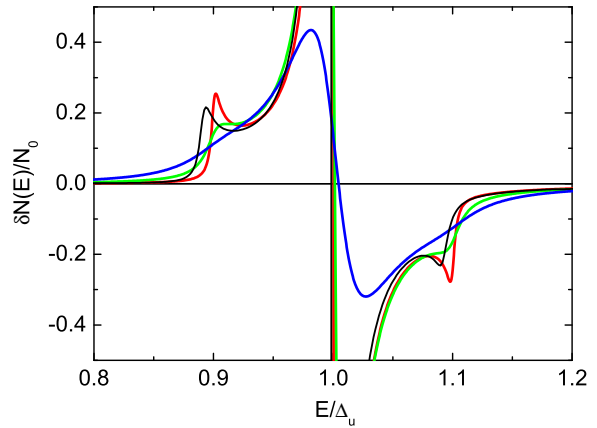} \caption{The change of normalized DOS of a superconductor, $\delta N\left(E\right)/N_{0}=Re[\delta\cos\theta\left(E\right)]$,
under the influence of an rf-drive with $\alpha/\Delta_{u}=10^{-3}$.
The black curve corresponds to the absence of relaxation, the red
curve corresponds to the value of the relaxation rate{} $\Gamma_{inel}/\Delta_{u}=0.003$,
the green curve corresponds to $\Gamma_{inel}/\Delta_{u}=0.01$, and
the black curve corresponds to $\Gamma_{inel}/\Delta_{u}=0.03$.}
\label{Figure2_DOS} 
\end{figure}

The change of the density of states $\delta N=N_{0}\mathrm{Re}[\delta\cos\theta]$,
with $\delta\cos\theta$ given by Eq.~\eqref{deltaG}. It
consists two terms, both proportional to the normalized field intensity: 

\noindent 
\begin{equation}
\delta N=\frac{\partial N}{\partial\alpha}\alpha-\frac{\partial N}{\partial\Delta}\frac{\pi}{\Delta_{u}}\alpha.\label{dDOS}
\end{equation}
The terms have different physical meaning. The first term is of the
main interest, because it describes qualitative modification of DOS
due to the embedded microvave field. Its magnitude is given by

\begin{equation}
.\begin{array}{c}
\frac{\partial N}{\partial\alpha}=N_{0}\mathrm{Re}\left[\frac{i\Delta_{u}^{2}\left\{ \left(E_{+}+i\Gamma_{inel}\right)+\left(E+i\Gamma_{inel}\right)\right\} }{\Xi_{+}\Xi^{3}}\right]+\\
+\left\{ E_{+}\rightarrow E_{-}\right\} 
\end{array}\label{dDOS_dalpha}
\end{equation}

Because of the factors $\Xi_{\pm}=\left\{ \left(E\pm\hbar\omega_{0}+i\Gamma_{inel}\right)^{2}-\Delta_{u}^{2}\right\} ^{1/2}$
in the denominator, it has features near the 'photon point' energies
$\Delta_{u}\mp\hbar\omega_{0}$. Near these energies, Eq.~\eqref{dDOS_dalpha}
can be approximated as

\begin{equation}
\frac{\partial N}{\partial\alpha}=N_{0}\frac{\Delta_{u}^{2}\left[\left(E_{\pm}-\Delta_{u}\right)+\left\{ \left(E_{\pm}-\Delta_{u}\right)^{2}+\Gamma_{inel}^{2}\right\} ^{1/2}\right]^{1/2}}{\left(2\hbar\omega_{0}\right)^{3/2}\left\{ \left(E_{\pm}-\Delta_{u}\right)^{2}+\Gamma_{inel}^{2}\right\} ^{1/2}}.\label{dDOS_dalpha-1}
\end{equation}
Its maximum scales as $2^{-3/2}\left(\Gamma_{inel}/\Delta_{u}\right)^{-1/2}\left(\hbar\omega_{0}/\Delta_{u}\right)^{-3/2}$,
and the width of the maximum is given by $\Gamma_{inel}$.

The second term of Eq.~\eqref{dDOS},

\begin{equation}
\frac{\partial N}{\partial\Delta}=-N_{0}\frac{\Delta_{u}\left(E+i\Gamma_{inel}\right)}{\Xi^{3}},\label{dDOS_dDelta}
\end{equation}

\noindent describes the shift of DOS due to the suppression
of the order parameter, $\Delta$, under influence of the microwave
field. It does not contain the photon-energy $\hbar\omega_{0}$.

The total change of DOS \eqref{dDOS} is presented in Fig.~\ref{Figure2_DOS},
for a fixed frequency $\hbar\omega_{0}=0.1\Delta_{u}$
and a fixed $\alpha=10^{-3}\Delta_{u}$. The change is normalized
to the normal state DOS. The inelastic collision-strength $\Gamma_{inel}/\Delta_{u}$
is varied from 0.003 (red), 0.01 (green), to 0.03
(blue). The smallest value almost coincides with the unperturbed
curve (black), calculated within the approach published
previously~\cite{semprl}. This indicates that the violation of the
condition$\alpha\ll\Gamma_{inel}$, which was needed to apply the
linear expansion in $\alpha$ near the peaks at $\Delta_{u}$ and
$\Delta_{u}\pm\hbar\omega_{0}$, does not affects significantly the
results for the spectral functions, confirming that this condition
is in practice not important. It is clear that the inelastic processes
reduce the visibility of the photon-structures in the density of states.
However, a quantitative analysis of the behavior of Eq.~\eqref{dDOS}
shows that the extremes in $\mathrm{Re}\delta\cos\theta\left(E\right)$
near $\Delta_{u}\pm\hbar\omega_{0}$ exist up to $\Gamma_{inel}/\hbar\omega_{0}\approx0.15$
(for $\hbar\omega_{0}\ll\Delta_{u}$). Hence, in principle the photon-steps
should be clearly discernible, provided a high enough accuracy can
be obtained in an experiment and sufficiently low temperatures are
used.

The most important quantity for kinetic induction detection and parametric
amplification is the non-linear kinetic inductance, as expressed below
in Eq.~\eqref{delta_Lk_Lk0_final}. It is a clear experimental signature
of how the microwave intensity gets embedded in the Cooper-pair condensate.
From a practical point of view that particular result is directly
usable in a model. Unfortunately, it is not very informative about
the influence of the microwave-field on the microscopic properties
of the superconductor and the dependence on the materials properties.
A much more critical test would be a direct measurement of the density
of states in the presence of microwaves.

\begin{figure}[h]
\centering \includegraphics[width=0.8\columnwidth]{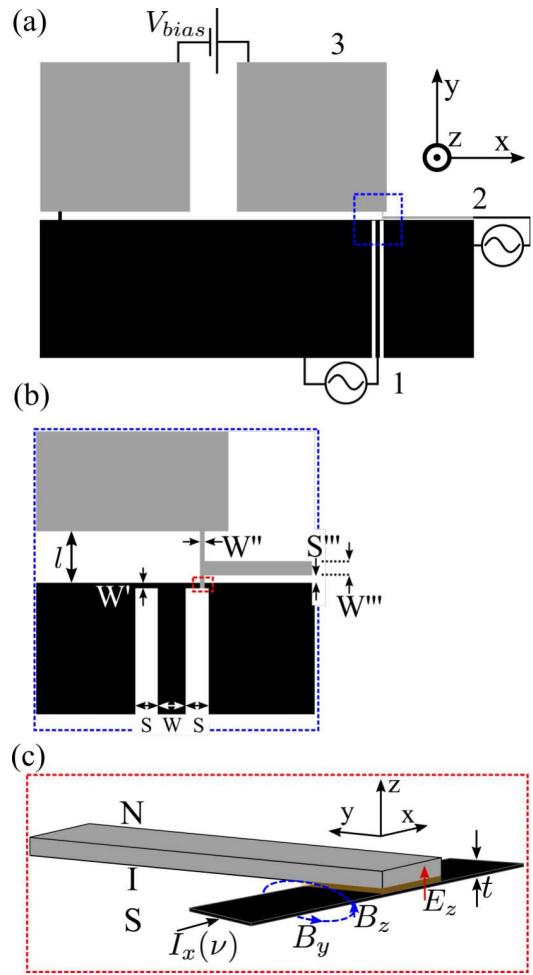}
\caption{(a) Proposed device to test the theoretical model. Black layers are
patterned in a superconductor whereas grey layers are patterned in
a normal metal. The circuit lies in the xy-plane of the indicated
coordinate system. (b) Blow-up of the region around the normal metal-insulator-superconductor
tunnel junction. (c) Side view of the tunnel junction (dark red) specifying
the created electric and magnetic fields by the rf current $I_{x}(\nu)$
at the driving frequency $\nu=\omega_{0}/2\pi$. The tunnel junction
is used to probe the coherent excitation of the (black) superconducting
wire beneath the tunnel junction by means of a density-of-states measurement
of the wire. The circuit sketch in (a) and (b) is drawn to scale.}
\label{fig01} 
\end{figure}

Here, we propose an experiment in which the rf-driven superconducting
properties are measured with a tunnel junction. It is well known that
tunnel junctions are very suitable to determine the density-of-states
as well as the Fermi distribution function of the superconductor.
However, the challenge is to design an experiment in which only one
of the electrodes is driven by the microwave field and not the other
electrode. In addition, one wants to avoid that in the measurement
by the tunnel-current the tunnel-process is modified by photon-assisted
tunneling (PAT) \cite{tie}. This problem has plagued early experiments
by Kommers and Clarke \cite{Kommers1977} and has led to some early
solutions by Horstman and Wolter \cite{hw1,wh1}. The experimental
challenge is to avoid or minimize an rf-field across the tunnel-barrier,
using the present-day fabrication technology and design tools. Fig.~\ref{fig01}(a)-(c)
shows our proposed experiment, which takes these considerations into
account.

Our proposed circuit can be divided into two parts which are shown
in black and grey in Fig.~\ref{fig01}. The black layer is a superconductor,
for instance aluminum, patterned as indicated in the figure. The grey
layer is made of a normal conducting metal, for instance copper. The
dashed blue box in Fig.~\ref{fig01}(a) indicates the region where
a normal metal-insulator-superconductor (NIS) tunnel junction is formed
between the black and grey metals. This region is shown in more detail
in Fig.~\ref{fig01}(b). The NIS tunnel junction is formed at the
overlap of the black superconducting wire with width $W'=1~\mu\mathrm{m}$,
thickness $t=20~\mathrm{nm}$, and the grey wire with width $W''=1~\mu\mathrm{m}$.
The latter normal metal wire has a total length of $l=21~\mu\mathrm{m}$
and acts as an inductance, just large enough, $L\approx15~\mathrm{pH}$,
to block the rf currents $I_{x}(\nu)$ to the NIS junction. Through
this we prevent that they propagate into the junction and couple into
the measurement circuitry attached to it. This signal blockage works
well in combination with an effectively shorted wire (black layer)
on which the junction is patterned, explained in more detail in the
following paragraph. At the same time the length of normal metal wire
that connects to the NIS junction is short enough to avoid a relevant
series resistance, which gets added to the overall junction tunnel
resistance. For a copper wire of the chosen dimensions with a common
thin film resistivity of $\rho_{0}=0.4~\mu\Omega$cm \cite{Boogaard2004}
one expects about $0.02~\Omega$ of series resistance. The NIS junction
tunnel resistance should, therefore, have a value much larger than
this series resistance, which is compatible with an opaque tunnel-barrier
to probe the superconducting properties. The NIS tunnel junction is
connected to three measurement terminals, labeled 1 to 3 in the figure.
They make it possible to probe the dc tunneling curve of the NIS junction,
while the states in the superconducting wire with width $W'$ underneath
the tunnel junction can be probed. Although in our device proposal
the NIS tunnel junction has an area of $1~\mu\mathrm{m}^{2}$, a smaller
tunnel junction will equally suffice to perform the experiment and
will only insignificantly modify the circuit functionality. Therefore,
our design is compatible with the established tri-layer and angle-evaporation
junction fabrication techniques that can realize junctions of different
sizes.

Terminal 1 is connected to a radio-frequency (rf) generator and is
used to excite a transversal electromagnetic (TEM) wave on a coplanar
waveguide (CPW) transmission line up to frequencies of 60 GHz. The
CPW is designed to have a characteristic impedance of $Z_{c}=50~\Omega$,
which we achieve by the CPW dimensions of $S=8~\mu\mathrm{m}$ and
$W=11~\mu\mathrm{m}$ on top of a $275~\mu\mathrm{m}$ thick silicon
wafer, ignoring a natural silicon oxide layer of about $1~\mathrm{nm}$.
The CPW is terminated as a short circuit by the superconducting wire
with width $W'$ at the position of the NIS junction. The short circuit
results in a maximum and homogeneous rf current $I_{x}(\omega_{0})$
in the wire and drives the superconducting ground state in the wire
at a particular frequency, $\omega_{0}$, and with a certain rf current
magnitude which can be adjusted at the rf generator. We find by modeling
our circuit in CST \cite{cst} that at the junction tunnel barrier
a magnetic and an electrical field is established due to the rf-driving,
as sketched schematically in Fig.~\ref{fig01}(c). We designed the
circuit in such a way to minimize primarily the electrical field component
$E_{z}$, established across the junction and which would lead to
unwanted PAT currents. If too large in magnitude, the PAT currents
would overwhelm the features due to the coherently excited density-of-states,
created on purpose by to rf-drive from terminal 1. For an rf-drive
power of -20~dBm at terminal 1, we find that the rf current which
goes through the superconducting wire at the position of the tunnel
junction will create a magnetic field of $B_{y},B_{z}<1$~G at the
tunnel barrier. Hence, it will only slightly disturb the superconducting
ground-state that we want to study. For the same drive power we expect
additionally the build-up of an electrical (stray) field $E_{z}$
across the junction which on average will amount to 4.5~V/m, leading
to a parasitic voltage drop of only 4.5~nV across the junction for
a tunnel barrier of 1~nm thickness. Dependent on the differential
resistance of the NIS junction under the rf-drive of the order of
several 100~$\Omega$, this will cause only a negligible parasitic
tunnel current. The magnetic and electrical field values are determined
for an excitation frequency of $\omega_{0}/2\pi=15$~GHz, but will
only slightly change for the other frequencies. Although not specified
in the figure, we envision to connect the rf generator through a circulator
or a directional coupler to terminal 1. This way we prevent the build-up
of a standing wave due to the reflection of the TEM wave at the wire
terminating the CPW.

A second rf generator can be connected to terminal 2 and could be
employed to excite a quasi-TEM wave on a coplanar strip (CPS) transmission
line which is connected from the right side to the NIS junction. One
part of the CPS transmission line connects to the S-part and the other
part connects to the N-part of the NIS junction, hence, an rf-current
is driven on purpose through the junction leading to a controlled
PAT current. This allows to disentangle possible PAT features which
might be introduced by exciting the circuit from terminal 1 and which
might disturb the density-of-states and distribution function measurements.
The coplanar strip transmission line has a characteristic impedance
equal to $Z_{c}=50~\Omega$, which we achieve by the dimensions $S'''=3~\mu\mathrm{m}$
and $W'''=7~\mu\mathrm{m}$. We suggest to use a normal metal for
the part of the CPS which connects to the N-part of the NIS junction
in order to prevent that the proximity effect modifies the density-of-states
in the NIS junction. Similar to the rf excitation from terminal 1,
we also suggest to connect the rf generator at terminal 2 through
a circulator or a directional coupler.

Finally, terminal 3 realizes the dc-bias or low-frequency part of
the circuit to voltage bias the NIS junction or to apply a low frequency
bias modulation for lock-in measurements of the differential resistance.
The latter measurement yields a convolution of the density-of-states
with the distribution function of the two NIS junction electrodes,
which are both unknown, but should be disentangled by a proper analysis
of the measurements obtained for different drive powers. For the same
reason we propose to use an asymmetric NIS junction for the deconvolution
procedure. Also, because of the applied voltage to the NIS junction,
we suggest to use DC-blocks at the terminals 1 and 2 to protect the
rf generators.

To fully characterize our device proposal, we need to quantify also
the isolation of the three terminals from each other when an rf excitation
is applied to them. We find in our circuit simulation reasonable isolation
values of $<-20~\mathrm{dB}$ for $S_{21},S_{31},S_{12}$ and $S_{32}$
and for the operation frequency band 2-60~GHz. Therefore we believe
that by using the currently available technology an evaluation of
the microscopic properties of a superconductor in the presence of
microwaves is feasible.

\section{Non-linear superconducting kinetic inductance}

Another quantity, which is important for microwave kinetic-inductance
detectors and parametric amplifiers, and determined by the change
in spectral properties and/or the distribution function, is the complex
conductivity $\sigma$ at frequency $\omega$. It is given by: 

\begin{equation}
\begin{aligned} & \sigma(\omega)=\frac{\sigma_{N}}{4\hbar\omega}\int dE\{\left(\cos\theta_{-}\mathrm{Re}\cos\theta+i\sin\theta_{-}\mathrm{Re}\left[i\sin\theta\right]\right)f_{L}-\\
 & -\left(\left(\cos\theta\right)^{*}\mathrm{Re}\cos\theta_{-}+\left(i\sin\theta\right)^{*}\mathrm{Re}\left[i\sin\theta_{-}\right]\right)f_{L-}\}.
\end{aligned}
\label{sigma}
\end{equation}

\noindent The imaginary part of the conductivity, measurable through
the kinetic inductance $L_{k}$, is given by the relationship: $L_{k}={1}/{\omega\mathrm{Im}\sigma}$.
Equation \eqref{sigma} is the generalization of the Mattis-Bardeen
relation \cite{mat} for the case of not only non-equilibrium distribution
functions, as was done by Catalani \textit{et al},\cite{nag}, but
also for changed spectral functions. For low frequencies $\hbar\omega\ll\Delta_{u}$,
the equation for the imaginary part of the conductivity \eqref{sigma}
reduces to

\begin{equation}
\mathrm{Im}[\sigma\left(\omega\ll\Delta/\hbar\right)]=\mathrm{Im}\sigma_{0}=-\frac{\sigma_{N}}{\hbar\omega}\int dE\mathrm{Im}\left[{\sin^{2}\theta}\right]f_{L}.\label{imsigma}
\end{equation}
The unperturbed value of $\mathrm{Im}\sigma\left(\omega\ll\Delta/\hbar\right)$
is given by

\noindent 
\begin{equation}
\mathrm{Im}\sigma_{0,u}=\sigma_{N}\frac{\Delta_{u}}{\hbar\omega}\pi,\label{imsigma0}
\end{equation}

\noindent which is a form of the above-mentioned well-known relation
between kinetic inductance and normal resistance \cite{Zmuidzinas2012,Tinkham}.



The small correction to the kinetic inductance at low frequencies,
$\hbar\omega\ll\Delta_{u}$, \emph{i.e.} the case of microwave radiation
on commonly used superconductors, is the sum of two terms:

\begin{equation}
\frac{\delta L_{k}}{L_{k,u}}=\frac{\delta_{\alpha}L_{k}}{L_{k,u}}+\frac{\delta_{\Delta}L_{k}}{L_{k,u}}.\label{delta_Lk_Lk0}
\end{equation}

\noindent The first term describes the change of the kinetic inductance
due to the change of the spectral and distribution functions under
the influence of rf-drive,

\begin{equation}
\frac{\delta_{\alpha}L_{k}}{L_{k,u}}=-\frac{1}{\mathrm{Im}\sigma_{0,u}}\frac{\partial\mathrm{Im}\sigma_{0}}{\partial\alpha}_{|\Delta=\Delta_{u}}\alpha,\label{delta_Lk_Lk1}
\end{equation}

whereas the second term describes the change of the kinetic inductance
due to change of the order parameter $\Delta$:

\begin{equation}
\frac{\delta_{\Delta}L_{k}}{L_{k,u}}=-\frac{1}{\mathrm{Im}\sigma_{0,u}}\frac{\partial\mathrm{Im}\sigma_{0}}{\partial\Delta}_{|\alpha=0}\delta\Delta.\label{delta_Lk_Lk2}
\end{equation}

Because of \eqref{imsigma0} the second term, Eq.~\eqref{delta_Lk_Lk2},
equals to ${\delta_{\Delta}L_{k}}/{L_{k,u}}=-{\delta\Delta}/{\Delta_{u}}$
and is given by \eqref{delta_Delta}. The first term, Eq.~\eqref{delta_Lk_Lk1},
is evaluated using the Eq.~\eqref{imsigma} for the imaginary part
of the conductivity:

\begin{equation}
\begin{aligned} & \frac{\delta_{\alpha}L_{k}}{L_{k,u}}=\\
 & =\left(\int dE\frac{\partial\mathrm{Im}{\sin^{2}\theta}}{\partial\alpha}f_{L,u}+\int dE\mathrm{Im}\left[{\sin^{2}\theta_{u}}\right]\frac{\partial f_{L}}{\partial\alpha}\right)_{|\Delta=\Delta_{u}}\alpha.
\end{aligned}
\label{d_alpha_Lk_Lk0}
\end{equation}

\noindent The second integral in this equation, which describes the
contribution due to the change of the distribution function, is negligible
for the same reason as the analogous contribution to $\delta\Delta$
in Eq.~\eqref{vardel}. The first integral in Eq.~\eqref{d_alpha_Lk_Lk0}
can be evaluated analytically in a way analogous to the one used for
Eq.~\eqref{delF}, by taking into account the Eqs.~\eqref{F0} and
\eqref{partF} and replacing $E$ to $iE$:

\begin{equation}
\begin{array}{c}
\left(\int_{0}^{\infty}dE\frac{\partial\mathrm{Im}{\sin^{2}\theta}}{\partial\alpha}f_{L,u}\right)_{|\Delta=\Delta_{u}}\cong\\
\cong-\int_{0}^{\infty}dy\frac{8y^{2}}{\left\{ y^{2}+1\right\} ^{5/2}}=-\frac{8}{3}
\end{array}.
\end{equation}

\noindent Substituting this in the first term of Eq.~\eqref{d_alpha_Lk_Lk0},
we obtain

\begin{equation}
\frac{\delta_{\alpha}L_{k}}{L_{k,u}}=\frac{16}{3\pi}\frac{\alpha}{\Delta_{u}},
\end{equation}

\noindent and, finally,

\begin{equation}
\frac{\delta L_{k}}{L_{k,u}}=\left(\frac{16}{3\pi}+\pi\right)\frac{\alpha}{\Delta_{u}}\simeq4.84\frac{\alpha}{\Delta_{u}}.\label{delta_Lk_Lk0_final}
\end{equation}
To facilitate comparison to Eq.~\eqref{NonlinearInductance}, we
rewrite Eq.~\eqref{delta_Lk_Lk0_final} in terms of $\left\langle I_{rf}^{2}\right\rangle /I_{*}^{2}$,
using Eq. \eqref{alpha_to_I}:

\begin{equation}
L_{k}(\left\langle I_{rf}^{2}\right\rangle )\approx L_{k}\left(0\right)[1+\left\langle I_{rf}^{2}\right\rangle /I_{*}^{2}]\label{NonlinearInductanceMicro}
\end{equation}
where $I_{*}=\sqrt{2}(\frac{16}{3\pi}+\pi)^{-1/2}\pi\Delta_{u}/eR_{\xi}\simeq2.02\Delta_{u}/eR_{\xi}\simeq2.69I_{c}$,
\textendash{} exactly as in the dc case.

This correction to the kinetic inductance, Eq.~\eqref{delta_Lk_Lk0_final},
as well as the correction to the order parameter ${\delta\Delta}/{\Delta_{u}}$
\eqref{delta_Delta}, agree with the values found in the previous
numerical calculation\cite{semprl}, despite the qualitative difference
between the unperturbed Green's functions, as well as between the
corrections to them in the presence of rf-drive $\delta\sin\theta$
and $\delta\cos\theta$. 
In our view this agreement has the following reason. The corrections
to $\Delta$ and $L_{k}$, as well as to other quantities which are
calculated as integrals of some spectral functions in infinite or
semi-infinite limits (and hence are not sensitive to the value of
$\Gamma_{inel}$), are expected to be the same if calculated in both
models. This is despite of the fact that the linearization procedure
presented in this article requires $\alpha\ll\Gamma_{inel}$, whereas
the derivations in Ref.~\cite{semprl} correspond to the opposite
limit: $\Gamma_{inel}\ll\alpha$. This indicates that introduction
of $\Gamma_{inel}$ in the Eq.~\eqref{theta_accurrent2_inel} can
be considered, formally, as a trick which allows the linearization
with respect to the ratio $\alpha/\Delta_{u}$. It shifts the poles
of the Green's functions away form the real axis and removes singularities,
which would render the linearization unfeasible.

There is one more (and more deep) consequence of the mentioned independence
of this type of integral quantities on the exact position of the poles.
In the final formulas, not only $\Gamma_{inel}$ but also $\hbar\omega_{0}$
do not matter, \emph{i.e.} one can safely replace $E\pm\hbar\omega_{0}$
by $E$. Noting that the same replacement (and $\Gamma_{inel}\rightarrow+0$)
in the retarded Usadel equation turns it into the form of the equation
for the dc case, with the depairing parameter $\Gamma=2\alpha$, one
sees that the corrections to these integral quantities should be equal
in the rf-case with the one in the dc-case. Actually, the results
\eqref{delta_Delta} and \eqref{delta_Lk_Lk0_final}, expressed in
terms of the root-mean-square value of the induced rf current ($\delta\Delta/\Delta_{u}=-0.088\left\langle I_{rf}^{2}\right\rangle /I_{c}^{2}$
and $\delta L_{k}/L_{k,u}=0.136\left\langle I_{rf}^{2}\right\rangle /I_{c}^{2}$),
exactly coincide with those for the dc-depairing theory \cite{ant},
with $\left\langle I_{rf}^{2}\right\rangle \rightarrow I_{dc}^{2}$.
Physically, this means that the time-averaged quantities are sensitive
only to the averaged kinetic energy contained in the supercurrent
(the condensate of Cooper pairs), but not to the frequency of its
oscillations. We want to stress that this equivalence between the
dc- and the rf-cases does not hold for the integral quantities, which
depend not only on spectral but also on the distribution functions,
like, for instance, the real part of the conductivity or the differential
conductance of an NIS tunnel-junction. The inequality $k_{B}T\ll\hbar\omega_{0}$
make these latter quantities sensitive to the exact position of the
poles at $E\pm\hbar\omega_{0}$.

We discuss briefly the applicability of these results to the analysis
of kinetic inductance traveling-wave parametric amplifiers \cite{eom}.
Typically, these devices, exploiting the nonlinearity of the kinetic
inductance induced by a strong supercurrent (pump), have to work under
the condition $\alpha\ll\hbar\omega_{0}$. The amplitude of the pump
supercurrent $I_{p}$ does not exceed $I_{c}/3$, hence $\left\langle I_{rf}^{2}\right\rangle /I_{c}^{2}\simeq1/20$
and the ratio $\alpha/\Delta_{u}\simeq0.028\left\langle I_{rf}^{2}\right\rangle /I_{c}^{2}\approx10^{-3}$.
For $\Delta_{u}/h\simeq300$ GHz (NbTiN or NbN) and $\omega_{0}/2\pi=1$
GHz, this yields $\alpha/\hbar\omega_{0}\simeq0.3$, and for $\omega_{0}/2\pi=10$
GHz yields $\alpha/\hbar\omega_{0}\simeq0.03$ (see, for instance
Refs.~\cite{eom},\cite{cha} and \cite{ada}). At the same time,
the simplified model used to describe operation of these devices assumes
that the kinetic inductance is altered as if the current were dc \cite{eom},
\emph{i.e.} is valid for the opposite case. Hence, the simplified
model has to be corrected or confirmed with the use of the theory
developed in Ref.~\cite{semprl} and the present one. To describe
the parametric interaction between two weak signals in a transmission
line, resonator, or lumped element, of which the kinetic inductance
is modulated by a strong pump, one has to know two quantities: the
nonlinear correction to the time-averaged admittance or kinetic inductance,
which is given by the formula \eqref{delta_Lk_Lk0_final}, and the
'cross-frequency' admittance $L_{cross}$, which is the coefficient
between the current at the frequency $\omega$ and the field at frequency
$2\omega_{0}-\omega$ (with $\omega_{0}$ the frequency of the pump).
To find $L_{cross}$, one needs the components of the spectral functions
oscillating at the frequencies $\pm2\omega_{0}$. This calculation
is beyond the scope of the present paper. Here, we just note that
in the case of low frequencies ($\hbar\omega_{0}\ll\alpha$), where
the dc-case equations are valid, the relationship $L_{cross}=\delta L_{k}/2$
holds\cite{eom}. For the rf-case, which is of interest here, we expect
that $L_{cross}$ depends only on $\alpha$ but not on $\omega_{0}$,
at least as long as $\omega_{0}\ll\Delta_{u}$. Hence, the answer
should be $L_{cross}=const\times\alpha/\Delta_{u}=const\times\delta L_{k}$,
which can differ from the prediction of the theory based on the dc-case
quantitatively, but not qualitatively.

\section{Conclusions}

In summary, we describe theoretically the influence of inelastic processes
on coherent excited states of a superconductor \cite{semprl}. We
considered the model, in which these processes are represented in
the relaxation approximation, which analogous to exchange of electrons
via tunneling to a normal reservoir \cite{melkop}. We have calculated
analytically the spectral functions as well as the nonequilibrium
distribution function in the presence of a monochromatic rf-drive.
We have demonstrated that when the conditions of the 'quantum mode
of depairing' are fulfilled, the change of the kinetic inductance
is determined primarily by the change of the spectral functions, and
not by the distribution function, which confirms the previously published
results\cite{semprl}.We have argued that our results are of a general
meaning, independent of a specific model for the inelastic relaxation.
We have discussed the implications for kinetic inductance traveling
wave parametric amplifiers. Finally, we have presented a full design
of an experiment to measure the predicted modification of DOS by embedded
microwave field, within reach of present day technology. 
\begin{acknowledgments}
We are grateful to M. Skvortsov, K. Tikhonov and P. J. de Visser for
stimulating and helpful discussions. We acknowledge financial support
from the Russian Scientific Foundation, Grant No. 17-72-30036. 
\end{acknowledgments}


\begin{thebibliography}{10}
\bibitem{key-1}We note a factor of 8 difference between our definition
of $\alpha$ and the one used in Ref.~\cite{TSK}. We also point
out that there is a misprint, the omitted factor 1/4, in the definition
of $\alpha$ in Ref.~\cite{semprl}.
\bibitem{endnote1} The condition of 'quantum mode of depairing' $\alpha/\Delta\ll\hbar\omega_{0}/\Delta$
can be rewritten as $\left\langle I_{rf}^{2}\right\rangle /I_{c}^{2}\ll0.028\hbar\omega_{0}/\Delta$,
with $\left\langle I_{rf}^{2}\right\rangle $ the mean-square amplitude
of the induced rf current and $I_{c}$ the depairing current. 
\end{thebibliography}
\end{document}